\pdfoutput=1
\documentclass[pt10,journal=jacsat,manuscript=article]{achemso}
\setkeys{acs}{articletitle = true}

\usepackage{units}
\usepackage{graphicx}
\usepackage{amsmath}
\usepackage{esint}
\usepackage{amssymb}   
\usepackage{achemso}
\usepackage[version=3]{mhchem}
\usepackage{titlesec}

\titleformat{\section}[display]{\bfseries}{}{0.0ex}{}[] 
\titleformat{\subsection}[runin]{\bfseries}{}{0.0ex}{}[] 
\titleformat{\subsubsection}[runin]{\itshape}{}{0.0ex}{}[] 

\makeatletter

\let\oldmaketitle\maketitle
\let\maketitle\relax

\title{How the Chemical Composition Alone Can Predict Vibrational Free Energies and Entropies of Solids}

\author{Fleur Legrain}
\email{fleur.legrain@cea.fr}
\affiliation[CEA, LITEN]
{CEA, LITEN, 17 Rue des Martyrs, 38054 Grenoble, France}
\author{Jes\'{u}s Carrete}
\affiliation[CEA, LITEN]
{CEA, LITEN, 17 Rue des Martyrs, 38054 Grenoble, France}
\altaffiliation{Current address: Institute of Materials Chemistry, TU Wien, A-1060 Vienna, Austria}
\author{Ambroise van Roekeghem}
\affiliation[CEA, LITEN]
{CEA, LITEN, 17 Rue des Martyrs, 38054 Grenoble, France}
\author{Stefano Curtarolo}
\affiliation[Stefano2]
{Dept. Mech. Eng. and Materials Science, Duke University, Durham, NC 27708, USA}
\author{Natalio Mingo}
\email{natalio.mingo@cea.fr}
\affiliation[CEA, LITEN]
{CEA, LITEN, 17 Rue des Martyrs, 38054 Grenoble, France}    

\makeatother

\usepackage[english]{babel}
\begin{document}
\oldmaketitle
\begin{abstract}
\noindent
Computing vibrational free energies ($F_{vib}$) and entropies ($S_{vib}$) has posed a long standing challenge to the high-throughput {\it ab initio} investigation of finite temperature properties of solids. Here we
use machine-learning techniques to efficiently predict $F_{vib}$ and $S_{vib}$ of crystalline compounds in the Inorganic Crystal Structure Database. By employing descriptors based simply on the chemical formula and using a training set of only 300 compounds, mean absolute errors of less than 0.04 meV/K/atom (15 meV/atom) are achieved  for $S_{vib}$ ($F_{vib}$), whose values are distributed within a range of $0.9$ meV/K/atom ($300$ meV/atom.) In addition, for training sets containing fewer than 2,000 compounds the chemical formula alone is shown to perform as well as, if not better than, four other more complex descriptors previously used in the literature. The accuracy and simplicity of the approach mean that it can be advantageously used for the fast screening of phase diagrams or chemical reactions at finite temperatures.
\end{abstract}

\section{Introduction}
The calculation of stability at high temperatures was identified four years ago as one of the major standing challenges for high-throughput (HT) {\it ab initio} approaches.\cite{Thehigh-throughputhighwaytocomputationalmaterialsdesign} 
Solving this problem is fundamental for the prediction of phase diagrams and of chemical reactions. HT phase diagrams have until now typically been calculated using the {\it ab initio} formation enthalpies at 0 K.\cite{Stabilityandelectronicpropertiesofnewinorganicperovskitesfromhigh-throughputabinitiocalculations,High-throughputabinitioanalysisoftheBiIn,ComprehensiveSearch} 
This means that in many cases the stable phases will not correspond to the ones at finite temperatures. With respect to chemical reactions, the common practice in HT has been to incorporate only the entropy of the gas phases, and to completely neglect the phonon vibrational contributions. This can lead to important errors in the estimated reaction pressures and temperatures.\cite{First-PrinciplesDeterminationofMulticomponent} 
The reason for this widespread neglect of the vibrational free energy is the high computational demand of phonon spectrum calculations. Whereas calculating the formation energy can be solved with calculations on a single unit cell, a vibrational entropy calculation requires computing the interatomic force constants of the solid in a large supercell, which increases the computational time by orders of magnitude. This has represented a bottleneck to the inclusion of the vibrational contribution in HT calculations of phase diagrams and chemical reactions. Solving it is a critical issue with a transforming impact for materials science.

This article shows that HT calculations of $S_{vib}$ and $F_{vib}$ of solids are feasible and can be carried out with satisfactory accuracy and less computational expense than previously thought. Furthermore, several Machine-Learning (ML) approaches are applied to this problem, and it is shown that the chemical composition alone can be efficiently used as a descriptor to rapidly predict vibrational properties, outperforming more sophisticated descriptors for small training sets. The predictive power of this method is validated by comparing predicted entropies of a dozen compounds with measured values from the National Institute of Standards and Technology (NIST) \cite{NIST}. The results presented here are a key step toward the HT screening of phase stability at finite temperatures, with important applications for the computational discovery of new materials including functional compounds such as hydrogen storage materials\cite{First-PrinciplesDeterminationofMulticomponent,PredictingReactionEquilibria} .

\section{Method}
\textit{Ab initio} HT computational methods are a powerful approach to identify
new application-specific materials.\cite{High-ThroughputComputationofThermalConductivity,UltralowThermalConductivityinFullHeuslerSemiconductors,ComputationalandexperimentalinvestigationofTmAgTe2,NanograinedHalf-HeuslerSemiconductors,High-throughputfirstprinciplessearchfornewferroelectrics,Thehigh-throughputhighwaytocomputationalmaterialsdesign,High-ThroughputComputationalScreeningofPerovskites,High-ThroughputScreeningofExtrinsicPointDefect,DissolvingthePeriodicTableinCubicZirconia,CarbonophosphatesANewFamilyofCathodeMaterials,DesigningMultielectronLithium-IonPhosphateCathodes,High-ThroughputDesignofNon-oxide,High-MobilityBismuth-based}
Instead of investigating materials one at a time, such methods use
algorithms to automate the calculations and analysis. However, the screening process can quickly require tremendous amounts of computational
resources because (i) ab initio calculations, usually performed using
density functional theory (DFT)-based methods, are computationally expensive,
(ii) up to hundreds of DFT runs per compound can be required to compute
some materials properties (e.g. anharmonic thermal conductivity\cite{UnravelingthedominantphononscatteringmechanisminthethermoelectriccompoundZrNiSn,Latticethermalconductivity}),
and (iii) the number of prospective candidates can easily climb into the hundreds of thousands. 

ML methods 
can provide a way 
to tackle this computational resources issue: instead of
running expensive DFT calculations for all prospective materials,
the materials properties are predicted very quickly using a ML model trained in advance. Using ML techniques
in such a way has allowed identifying materials with targeted properties\cite{MachineLearninginMaterialsScience,MaterialsCartographyRepresentingandMining,FromOrganizedHigh-ThroughputDatatoPhenomenologicalTheory},
such as compounds with unprecedentedly low thermal conductivity\cite{PhysRevX.4.011019,PredictionofLow-Thermal-ConductivityCompounds},
NiTi-based shape memory alloys with exceptionally low thermal hysteresis\cite{Acceleratedsearchformaterialswith},
and organic polymers with remarkably high band gap and dielectric
constant\cite{Rationaldesignofallorganicpolymerdielectrics}. 

ML methods build a model (which can be seen
as a function) that transforms inputs (also called descriptors, characterizing
the materials) into outputs (usually a materials property such as
thermal conductivity or dielectric constant) that should be as close
as possible to the targets (the actual values
of the material's property). ML methods operate in three
stages: (i) the learning phase in which the model is trained by minimizing
the differences between outputs and targets for a set of compounds
(ii) the test phase in which the model is tested by assessing the differences
between outputs and targets for a second set of compounds (different
from the training set) (iii) the prediction phase in which the model
is used to effectively predict the targets (unknown) of other compounds.

\subsection{Data sets.}

The data set consists of the Inorganic Crystal Structure Database (ICSD) section of the aflow.org repositories (52,671 compounds)\cite{AFLOWAnautomaticframework,Aforexchangingmaterialsdata,AFLOW104}
from which the duplicates are removed (reducing it to 25,705 compounds) as well as
the compounds containing the noble gases Ne, Ar, Kr, and Xe, the elements
from Cm (Z = 96) to Cn (Z = 112), and Tb, At, Rn, Fr (further reducing it to 25,075
compounds). High-throughput calculations are used to compute the phonon frequencies of about 600 compounds randomly selected among this set of 25,075
materials. Out of the 423 phonon calculations that ran smoothly, the 131 calculations which resulted in imaginary phonon frequencies are discarded, with a tolerance of $2$ rad/ps for the first three frequencies. The data set employed in the following to predict the vibrational properties consists of the 292 remaining calculations. 



\subsection{Descriptors.}
The set of descriptors is one of the main components that determine the performance of a ML method. This set should satisfy a few constraints: it should contain the same number
of features (or components) for all materials, and for instance not
depend on the number of atoms or atom types; it is also well accepted
that good sets of descriptors should be invariant under rotation and translation
of the materials structure as well as under permutation of atoms---so
that every material is characterized by a unique vector.
This means that, unless one restricts the materials of study to a
specific class (such as Heuslers\cite{NanograinedHalf-HeuslerSemiconductors,High-ThroughputMachine-Learning-Driven}
or ternary oxides\cite{FindingNatureMissingTernaryOxideCompounds}),
a lot of thought must be put into the design of appropriate descriptors. Many different
types of descriptors complying with such constraints have been proposed
in the literature. Some of them are structural: a non-exhaustive list
includes Coulomb matrices\cite{FastandAccurateModeling}, bags of
bonds\cite{MachineLearningPredictionsofMolecularProperties}, pair
correlation functions\cite{Howtorepresentcrystalstructuresformachinelearning}, graph-theoretic variables\cite{GraphTheoryMeets}. Other descriptors are physical, such as transformations (mean, standard deviation\dots) of properties (masses, radii\dots) of the atoms of the material\cite{Representationofcompoundsformachine-learningpredictionofphysicalproperties,UniversalFragmentDescriptorsforPredictingElectronicPropertiesofInorganicCrystals,Three-ParameterCrystal-StructurePrediction}. 

In this article, a set of descriptors based exclusively on the chemical composition is first employed. In a second step, the performance of four other more complex descriptors is explored: despite the popularity of ML techniques
in materials modeling very few studies have compared
different descriptors on the same basis.  
The four additional sets are based
on the pair correlation functions, O'Keeffe's description of solid angles, bispectrum
components, and the properties of the atoms of the material.

\subsubsection{Chemical composition.}

The first set of descriptors contains exclusively the chemical
composition information of the compounds. The vector of descriptors
has 87 components, each component $b_{j}$ being the fractional
composition of the compound in the element type $Z_{j}$ ($Z_{1}$
is H, $Z_{2}$ is He\dots ). For instance, \ce{Mg2Si} is described
by $(0,0,\ldots,0,\nicefrac{2}{3},0,\nicefrac{1}{3},0,\ldots,0)$. Seko and collaborators employed similar descriptors for the prediction of low-thermal-conductivity compounds, except that they use binary digits to represent the presence of chemical elements and they do not account for the fractional composition.\cite{PredictionofLow-Thermal-ConductivityCompounds}

\subsubsection{Elemental properties of the atoms.}

Twelve elemental properties of the atoms are considered:
atomic number $Z$, mass $m$, radius $r$, column number $col$,
row number $row$, electronegativity $\chi$, Pettifor scale $Ps$,
Pettifor number $Pn$, and number of s, p, d, f valence electrons
(represented by $s$, $p$, $d$, $f$). For all compounds the elements $x_{i,j}$ are generated, they give the value of the $i$th elemental property
for the $j$th atom of the unit cell. Five descriptors are defined for each of the 12 elemental properties: the mean, minimum, and maximum of the property
over the atoms, the value of the property for the most abundant chemical element
(or the mean of the property over the most abundant chemical elements if two or more chemical elements are equally dominant), and $\frac{2}{N}\sum_{j,k=1}^{N}\left|x_{i,j}-x_{i,k}\right|$
($N$ being the number of atoms), respectively denoted as \textit{mean},
\textit{min}, \textit{max}, \textit{ab}, and \textit{var}, and resulting in a set of 60 descriptors.

\subsubsection{Bart\'{o}k-Partay's bispectrum components.}

Bart\'{o}k-Partay et al. have developed a type of interatomic potentials, the Gaussian Approximation Potentials (GAP), based on the bispectrum components of the atoms.\cite{GaussianApproximationPotentialsTheAccuracy,SNAP} The bispectrum components, defined for each atom of the unit cell, are based on the local atomic density surrounding the atom and are meant to describe its atomic environment while being invariant
under translation, rotation and permutation of atoms. The bispectrum components are generated using the LAMMPS code \cite{FastParallelAlgorithms} and a radius cutoff of 6 $\mathrm{\mathring{A}}$. The band limit $j_{max}$ is set to 5 and only the diagonal components are considered, resulting in 52 components per atom. This gives for
all compounds the elements $b_{i,j}$, representing the $i^{th}$ bispectrum
component of the $j^{th}$ atom of the unit cell. The bispectrum
components are then averaged over the atoms of the unit cell of the same chemical type
to form the matrix $B'$ of size \textit{number of bispectrum components}
$\times$ \textit{number of different elements}, i.e. $52\times 87$, where the
element $b'_{i,j}$ represents the mean of the $i^{th}$ bispectrum component
over all atoms of chemical type $Z_{j}$ ($Z_{1}$ is H, $Z_{2}$
is He\dots ). The vector of descriptors employed consists of the 4524
$b'_{i,j}$ elements.

\subsubsection{O'Keeffe's solid angles.}

Inspired by O'Keeffe's definition of coordination numbers\cite{Aproposedrigorousdefinitionofcoordinationnumber},
a set of descriptors is based on the solid angles subtended by the
faces of the Voronoi polyhedron centered at each atom. The solid angles $\Omega_{i,j}$, centered at each atom $i$, represent the solid
angle between the atom $i$ and one of its neighbors $j$ (among its set
of neighbors $x_{i}$). The derived elements $o_{i,j}$
are computed as
: $o_{i,j}=\frac{1}{N_{i}}\sum_{k,l=0}^{N}\delta_{i,k}\delta_{j,l}\Omega_{k,l}$;
$N_{i}$ is the number of atoms of chemical type $Z_{i}$, and $\delta_{i,k}$
($\delta_{j,l}$) is equal to 1 if the $k^{th}$ ($l^{th}$) atom is of chemical
type $Z_{i}$ ($Z_{j}$) and 0 otherwise. This means that for each
atom of the unit cell the solid angles are computed for all its neighbors,
and summed up for neighbors of same chemical type $Z_{j}$. The solid-angle sums are then averaged over the atoms of the unit cell which
are of the same chemical type $Z_{i}$. The vector of descriptors
employed consists of the $87\times 87$ (7,569) $o_{i,j}$ elements.

\subsubsection{Pair correlation functions.}

Another set of descriptors uses the partial radial distribution function
representation that considers the distribution of pairwise distances
$d_{ab}$ between two atom types, $a$ and $b$, as described by Sch\"{u}tt
et al.\cite{Howtorepresentcrystalstructuresformachinelearning}
The functions are splitted into 200 bins, each bin spanning 0.1 $\mathrm{\mathring{A}}$,
which leads to a set of $200\times 87\times 87$ (1,513,800) descriptors.

\section{Results and discussion}

\subsection{$\Gamma$-only versus full Brillouin zone calculations.}

The vibrational entropies $S_{vib}$ and free energies
$F_{vib}$ can be computed at different temperatures based on the
phonon density of states. Such phonon density of states can be obtained with first-principles calculations,
using a rather dense phonon wave-vector \textit{q}-point grid or a large supercell. However,
because of computational resources, it is desirable to use a phonon density of states computed
using single unit cells and a limited number of \textit{q}-points. The most extreme simplification consists in only using the $\vec q = \vec 0$ (i.e. $\Gamma$ point) phonon frequencies.

It is possible to assess how the vibrational properties calculated using the
phonon frequencies at $\Gamma$  (i.e. coarser but much cheaper) compare with those
calculated using the full phonon spectra (i.e. more accurate but also
much more expensive). The data provided by Togo and collaborators\cite{phonopy,CommentaryTheMaterialsProject,PythonMaterialsGenomics,TheMaterialsApplicationProgrammingInterface}, consisting of 207 compounds, are used for this.
Because the acoustic frequencies $f_{ac}$, which are zero at $\Gamma$,
are generally non-zero at other points, one must use a non-zero representative of the acoustic frequencies. Here a fractional part of the average optical frequencies is used: $f_{ac}=\nicefrac{1}{2} <f_{opt}>$, where $f_{opt}$
represent the optical frequencies. The value $\nicefrac{1}{2}$ is taken as it is the value that gives best overall results.

\begin{figure}
\includegraphics[width=0.99\linewidth]{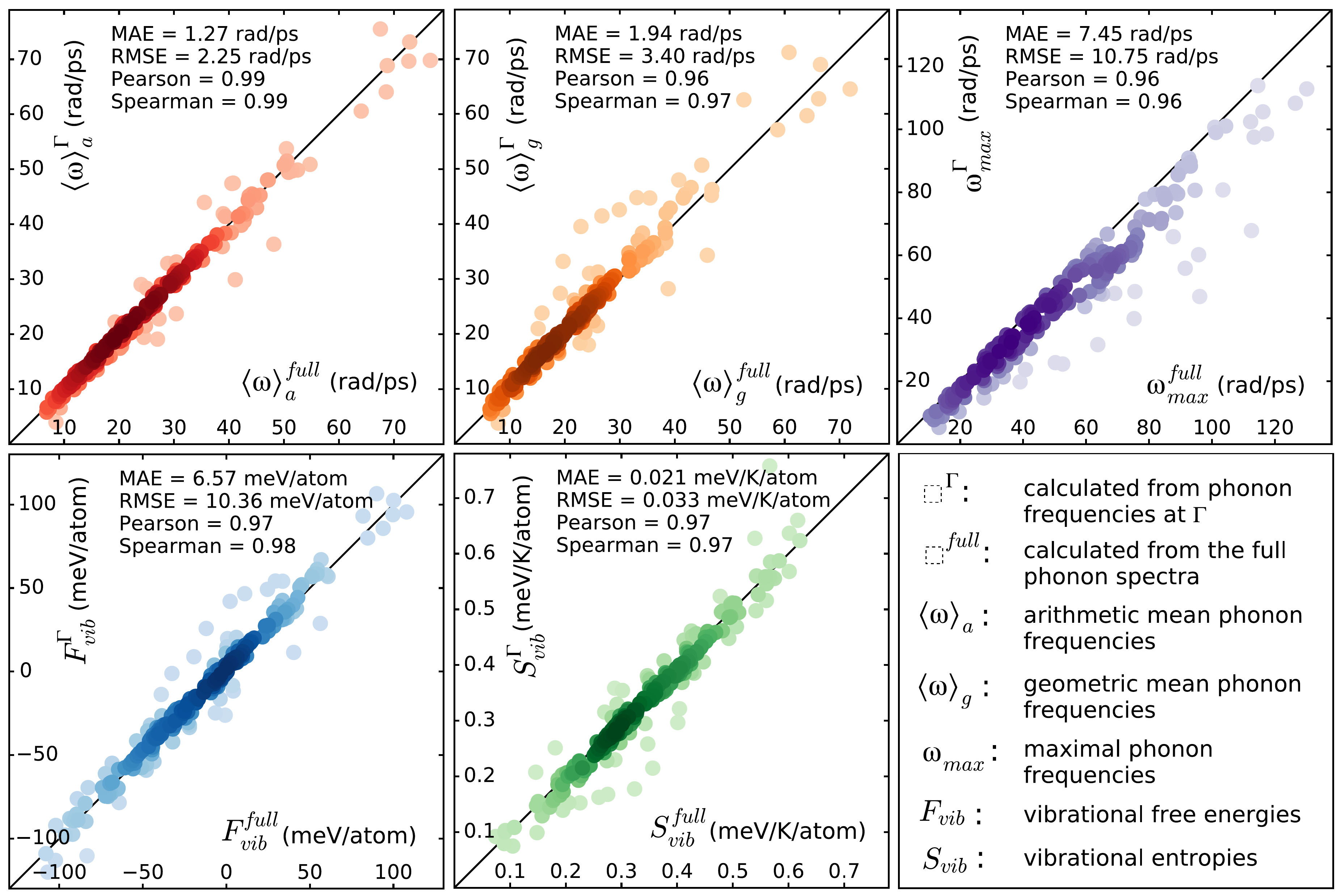}

\caption{Vibrational properties calculated from the phonon frequencies at $\Gamma$ \textit{vs.} those
calculated from the full phonon spectra. For the vibrational entropies and free energies, the temperature considered is 300 K. The mean absolute errors (MAE), root
mean squared errors (RMSE), and Pearson and Spearman correlations
are given.}

\end{figure} 

Figure 1 
displays 
the values of all the vibrational properties considered, calculated from the phonon frequencies at $\Gamma$, against the more accurate values calculated from the
full phonon density of states. On the plots, the mean absolute errors (MAE),
root mean squared errors (RMSE), Pearson and Spearman correlations are provided.
For all properties, the Spearman and Pearson correlations are higher
than 0.95, and for the maximal phonon frequencies, which are the least
well predicted properties, the MAE (7.45 rad/ps) represents about 6\%
of the range of the maximal frequencies (about 130 rad/ps). This shows
that the phonon frequencies at $\Gamma$ already result in well approximated values
of the vibrational properties. Because this approximation allows massive
computational resource savings, it is used to calculate the vibrational properties of the 292 compounds---together with the acoustic
frequencies as defined above.

\subsection{How well are vibrational properties predicted based solely on chemical composition?}

The phonon frequencies at $\Gamma$ are computed for randomly selected materials among
our data set of 25,075 compounds. After discarding erroneous calculations 292 sets of phonon frequencies are obtained. They are used to calculate the vibrational entropies $S_{vib}$ and free energies $F_{vib}$ per atom
at 300 K, the maximal phonon frequencies $\omega_{max}$, and the arithmetic $\langle \omega \rangle_{a}$
and geometric $\langle \omega \rangle_{g}$ means of phonon frequencies. The
performance of machine learning for the prediction of the vibrational
properties is then studied using the chemical compositions as descriptors. The machine learning algorithms are described in the Computational details section.

To assess the performance of the ML method, the k-fold cross validation
technique is employed with $k=5...14$ and the performance is averaged over all (i.e. 10) cross
validations. The corresponding MAE, RMSE, Pearson and Spearman correlations
are given in Figure 2, next to the plots showing the predicted \textit{vs.}
computed vibrational properties as obtained with the 14-fold cross
validation. Given the small training set (fewer than 300 compounds)
and the simplicity of the descriptors (containing only the chemical
composition), the performance of the ML approach is
impressive: the Pearson and Spearman correlations are superior to
0.9 for all properties, and the MAE is less than 6 rad/ps for the average phonon frequencies (for a range
of 250 rad/ps) and is less than 15 meV/atom for the vibrational free energies (for a range
of 300 meV/atom).

\begin{figure}
\includegraphics[width=0.99\linewidth]{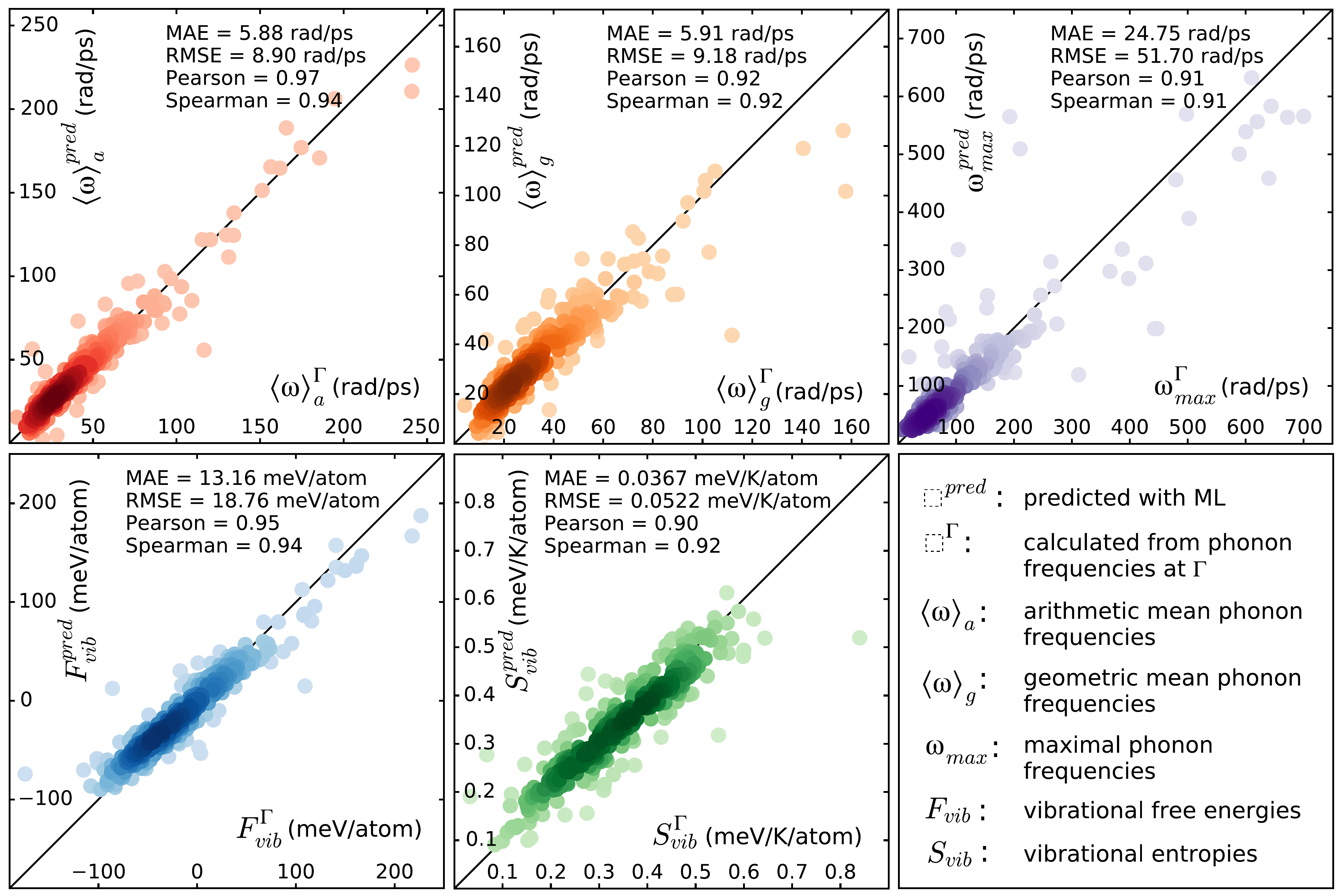}

\caption{Predicted \textit{vs.} computed vibrational properties. When computed, the phonon frequencies at $\Gamma$ are used. For the vibrational entropies and free energies, the temperature considered is 300 K. The MAE, RMSE, Pearson
and Spearman correlations are given for each property. }

\end{figure}

In a second step, the performance of the method is assessed by comparing
directly the vibrational entropy predictions with experimental values taken from
NIST\cite{NIST}. Compounds relevant for hydrogen storage are selected as vibrational entropies
can play a role in the stability of such compounds but they are often neglected. The materials considered are all the crystalline solids of Ref.~5
and of Table 3 of Ref.~7 for which
the entropy is available in NIST\cite{NIST}. Figure 3 shows the plot of the vibrational entropies as predicted with the ML model (trained on
the 292 compounds) against the ones measured experimentally. The agreement
between the predictions and the experiments is remarkable given the
simplicity of the method and the approximations done. This shows that the approach can effectively be employed to rapidly predict  the vibrational properties of crystalline compounds.

\begin{figure}
\includegraphics[width=0.5\linewidth]{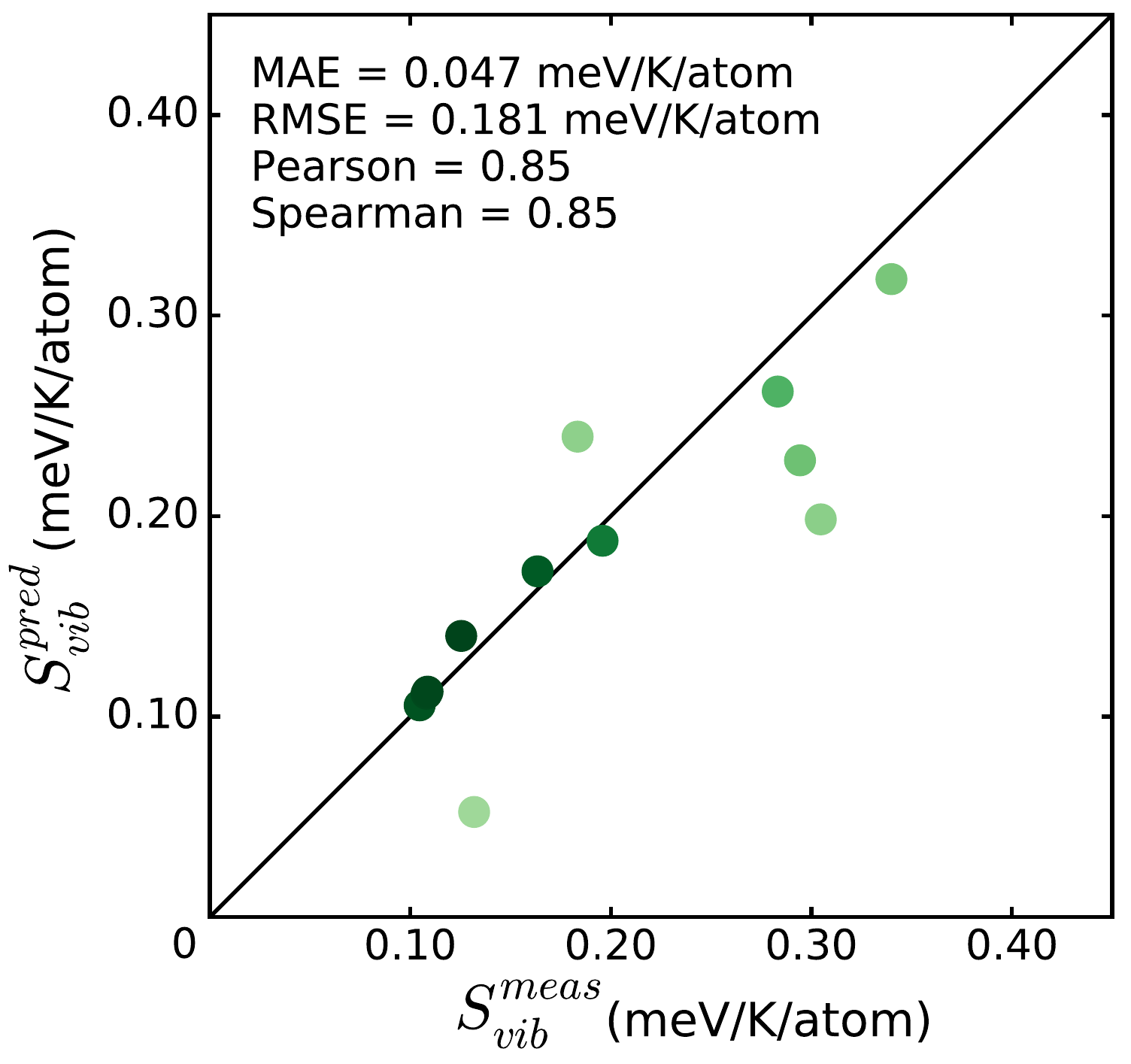}

\caption{Predicted \textit{vs.} measured vibrational entropies at 300 K for a dozen compounds arbitrarily chosen---the materials of Ref 5
and of Table 3 of Ref 7 for which the vibrational entropies are avalaible in NIST\cite{NIST}.}

\end{figure}

\subsection{What descriptors yield the best prediction of vibrational properties?}

So far the set of descriptors employed only contained the chemical composition
information. In particular, the inputs hold no information regarding
the structure of the compound or the physical properties of the atoms.
Now, in addition to the chemical composition, four additional
sets of descriptors are explored. They are based respectively on the pair correlation functions,
properties of the atoms, bispectrum components, and O'Keeffe's
solid angles. 

\begin{figure}
\includegraphics[width=0.6\linewidth]{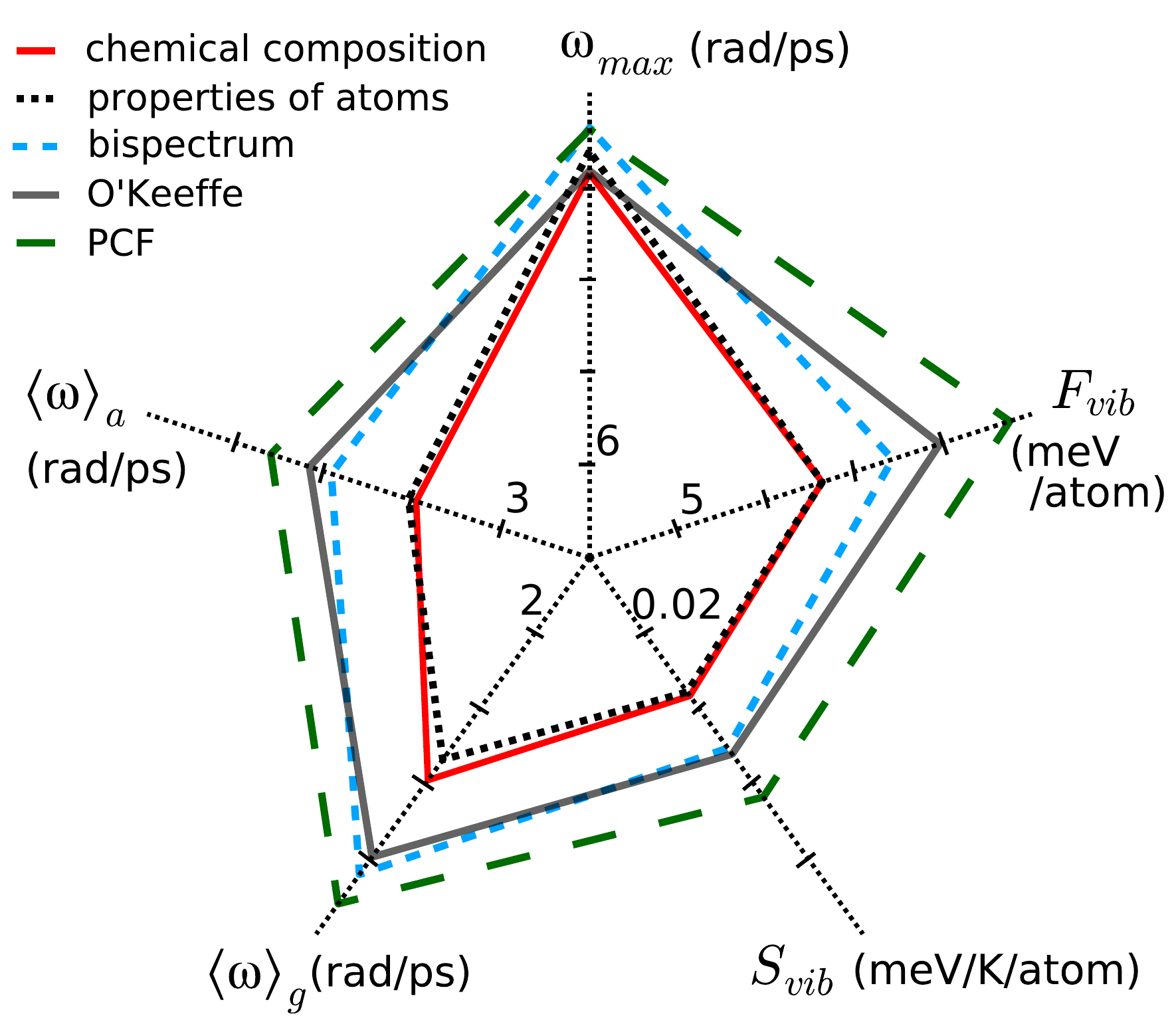}

\caption{MAE for the predictions of the five vibrational properties for the
descriptors based on the chemical composition, pair correlation functions
(PCF), O'Keeffe's solid angles, bispectrum components,
and elemental properties of atoms.}

\end{figure}

The performances of the different descriptors are evaluated using the MAE, shown
on Figure 4. The smaller the MAE, the more
accurate the prediction. The results show that the two most competitive
sets of descriptors are those based on the chemical
composition and on the elemental properties of the atoms. The three
other types of descriptors (based on the pair correlation functions,
O'Keeffe's solid angles, and bispectrum
components) do not perform as well. Interestingly,
the descriptors tend to work better when they contain fewer components:
the descriptors based on the pair correlation functions (with $200\times 87\times 87$
components) are overall the least effective set of descriptors, the
descriptors based on the O'Keeffe's
solid angles (with $87\times 87$ components) and on the bispectrum components
(with $52\times 87$ components) have a similar and intermediate overall
performance, while the most successful descriptors (based on the chemical
composition and on the elemental properties of the atoms) contain
respectively 87 and 60 components.


Figure 5 shows the correlogram between the vibrational entropies
and the properties of the 
elements in the compounds.
For each property of the atoms (atomic number, mass, radius, \dots ),
only the component most correlated with $S_{vib}$ (out
of \textit{mean}, \textit{min}, \textit{max}, \textit{ab}, and \textit{var}) is presented.
The
property 
most correlated with the vibrational entropies is the mean of the rows of the atoms $row_{mean}$. 
However,
the performances obtained when using this descriptor only are not nearly as good as using all the descriptors:
0.055 meV/K/atom (MAE), 0.071 meV/K/atom (RMSE), 0.81 (Pearson), and 0.80 (Spearman), \textit{vs.} 0.037 meV/K/atom (MAE), 0.051 meV/K/atom (RMSE), 0.91 (Pearson), and 0.92 (Spearman) when using the full set of descriptors. 
This shows that simple intuition cannot achieve the same result as the one obtained with machine learning. Machine learning affords results beyond what classical rules of thumb can provide.



\begin{figure}
\includegraphics[width=0.7\linewidth]{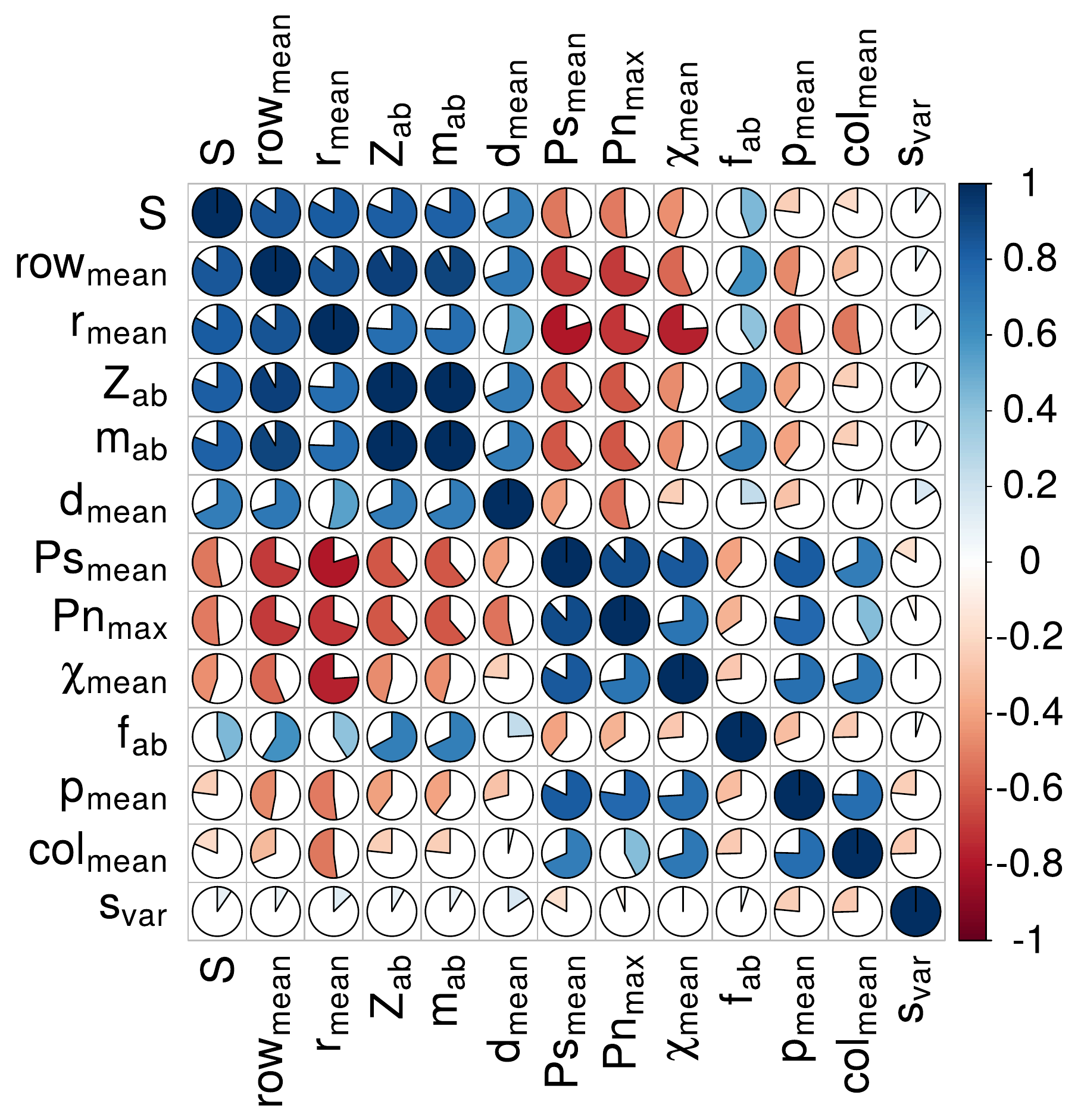}

\caption{Correlogram between the vibrational entropies per atom at 300 K and the elemental properties of the
atoms.}

\end{figure}


It is also useful to study the performance of these five different sets of descriptors for predicting properties other than the vibrational ones. The metallic or insulator character
of the material, provided
in the aflow.org repository, is particularly well suited for this. The advantage is that now the performance of the different descriptors can be evaluated as a function of the size of the training set,
which is much larger than the training set available for vibrational properties. For the prediction
of the metallic \textit{vs.} insulator (M/I) character, the whole set of 25,075 compounds
is considered. The M/I character of the materials is based on the
band gaps provided in the aflow.org repository. We are aware of the limitations of DFT to compute material band gaps, our focus is not to predict the M/I character of additional compounds but to study the predictive capabilities of the model.

\begin{figure}
\includegraphics[width=0.4\linewidth]{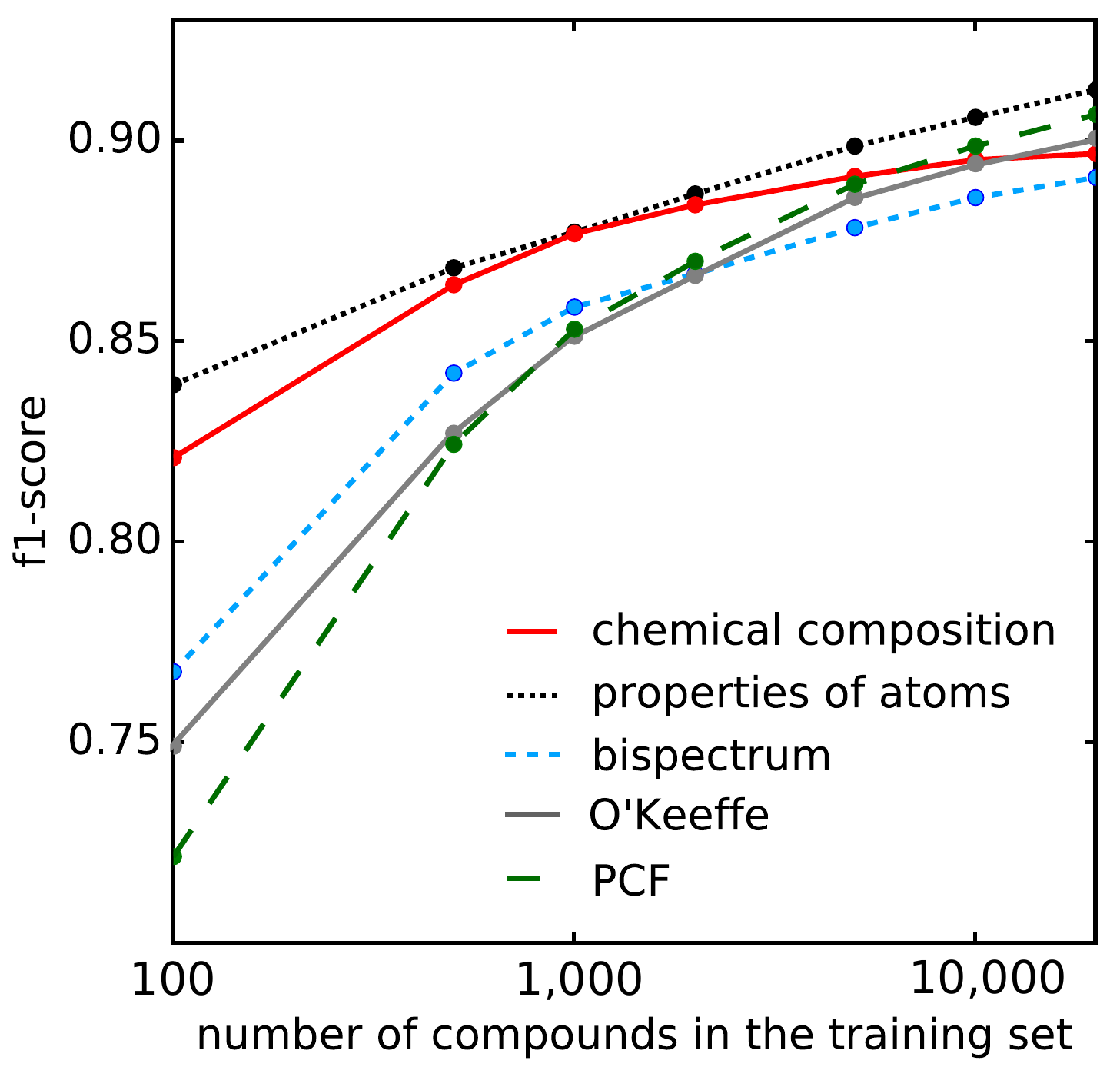}

\caption{The f1-score for the prediction of the metallic / insulator character
is plotted as a function of the training set size in a range of 100 to 20,000 components with a logarithmic scale.}

\end{figure}

Figure 6 shows the performances of the descriptors
against the size of the training set. The f1-score is used as indicator
of performance. The f1-score is defined for each class (metal and insulator) as the harmonic mean of precision and recall (times 2 to scale the score to 1): \textit{f1-score} $= 2 \times (recall \times precision)/(recall + precision)$. The global f1-score is calculated as a weighted average of the f1-score of each class. The larger the f1-score, the better the descriptors. The
results are consistent with our above discussed findings regarding the vibrational
properties: when the training set contains fewer than 2,000 compounds,
the descriptors based on the chemical composition and on the elemental
properties of atoms are the most powerful sets of descriptors, achieving similar performance.
However, as the size of the training set is increased, the sets of
descriptors based on the properties of the atoms, on the pair correlation functions, and on the O'Keeffe's
solid angles surpass the one based on the chemical composition.

\section{Conclusions}

Three major conclusions emerge from all the results above.

First, machine learning makes it possible to efficiently predict the vibrational entropies and free energies of crystalline compounds. Using a set of descriptors simply based on the compound's chemical formula, and a training set of 292 compounds in the ICSD, a MAE of 0.037 meV/K/atom (13.2 meV/atom) is achieved for the prediction of vibrational entropies (free energies), whose values range from 0 to 0.9 meV/K/atom (from -100 to 200 meV/atom). Excellent performance is also demonstrated by directly comparing predicted entropies with measured values from NIST for a dozen compounds relevant for hydrogen storage . The obtained MAE of 0.047 meV/K, and Pearson and Spearman correlations of 0.85, mean that the predicted entropies are good estimates of the measured values. This approach therefore allows for the rapid calculation of vibrational free energies and entropies, representing a key step toward the study of materials stability at finite temperatures using HT methods.  

A second conclusion is that the vibrational properties computed from the phonon frequencies at $\Gamma$
are already good approximations of the values calculated from the full phonon density of states. This is an important piece of information
in situations where the computational resources to compute the full
phonon density of states are not available.

The third conclusion is that, 
for
small
training sets, descriptors based on the chemical composition and
on the elemental properties of atoms of the material  
perform best.
It is only for much larger training sets, from
about a few thousands of compounds, that the set of descriptors based on the chemical composition is outperformed by some
of the more elaborate descriptors, like those based on the bispectrum, pair correlation functions, or O'Keeffe's
solid angles.

\section{Computational details}
The phonon frequencies of 292 randomly selected compounds are computed using
DFT\cite{InhomogeneousElectronGas,PhysRev.140.A1133}
as implemented in the Vienna Ab initio Package (VASP)\cite{Efficiencyofab-initiototalenergy}.
The projector augmented wave (PAW) method is employed to deal with the core and valence electrons\cite{Fromultrasoftpseudopotentials}. The specific choice of PAW datasets follows aflow.org's\cite{AFLOWAnautomaticframework,Aforexchangingmaterialsdata,AFLOW104} recommendations,
and the default cutoffs are used for the plane wave basis. The phonon frequencies are computed at $\Gamma$
using density functional perturbation theory\cite{DFTP}. From the phonon frequencies $\omega$ it is possible to obtain the vibrational entropies ($S_{vib}$) and free energies ($F_{vib}$), as well as the maximal phonon frequencies ($\omega_{max}$), and the arithmetic and geometric means ($\langle \omega \rangle_{a}$ and $\langle \omega \rangle_{g}$) of phonon frequencies. The vibrational entropies and free energies are computed as\cite{Landau}:
\newline
\begin{equation}
S_{vib}=\frac{3}{N_{freq}}\sum_{\omega}\left(\frac{\frac{\hbar\omega}{T}\exp\left(\frac{-\hbar\omega}{k_{B}T}\right)}{1-\exp\left(\frac{-\hbar\omega}{k_{B}T}\right)}
-k_{B}\ln{\left(1-\exp\left(\frac{-\hbar\omega}{k_{B}T}\right)\right)}\right)
\end{equation}

\begin{equation}
F_{vib}=E_{vib}-TS_{vib} 
=-\frac{3}{N_{freq}}\sum_{\omega}\left(\frac{\hbar\omega}{2}+k_{B}T\ln{n_\omega}\right)
\end{equation}
\noindent
In the equations, $N_{freq}$ is the number of phonon frequencies and $n_\omega$ is the Bose-Einstein factor.

To predict the vibrational properties and the metallic / insulator character of the materials, two different types of ML algorithm are systematically employed: random forests and non-linear support
vector machines. The number of trees is set to 500 for the random
forests. It is checked that an increase in the number of trees does
not result in better performance. For non-linear support vector machines, the radial basis function kernel is used and the $\gamma$ and $C$ coefficients are optimized for each different descriptors-properties system. For the prediction of the metallic
/ insulator character for which different training sets are considered, the $\gamma$ and $C$ coefficients are optimized for a training
set of 1,000 compounds and the optimized values are used for the
other training sets. Only the best performance is presented, which is sometimes obtained using random forests,
and at other times using non-linear support vector machines. The performance
is assessed by calculating the mean absolute errors between the predictions and the targets for a set
of compounds not included in the training set. For the prediction
of the metallic / insulator character of the materials for which there is a data set of 25,075 compounds, the model is trained with the training
set (containing $X$ compounds, $X$ being in the range 100-20,000) and
the performance of the model is assessed using the remaining data (i.e.
a set of 25,075 - $X$ compounds). The process is repeated with 10 different
(and randomly selected) training sets, and the average performance is presented.
For the prediction of the vibrational properties, for which the data
set contains 292 compounds, 10 k-fold cross validations
(k = 5...14) are performed and the performances obtained are averaged.

\section{Acknowledgements}
The work is supported by M-era.net through the ICETS project (DFG: MA 5487/4-1). We also acknowledge support from ANR through the Carnot MAPPE project.

\begin{doublespacing}

\bibliographystyle{achemso}
\bibliography{bib}

\end{doublespacing}

\end{document}